\documentclass[a4paper,10pt,twoside]{cpc-hepnp}

\usepackage{multicol}
\usepackage{graphicx}
\usepackage{booktabs}
\usepackage{amssymb,bm,bbm,amscd}
\usepackage[tbtags]{amsmath}
\usepackage{lastpage}
\usepackage{CJK}

\begin{document}

\fancyhead[c]{\small Chinese Physics C~~~Vol. XX, No. X (201X)
XXXXXX} \fancyfoot[C]{\small 010201-\thepage}

\footnotetext[0]{Received 15 May 2014}

\title{Is $J^{PC}=3^{-+}$ molecule possible?\thanks{Supported by National Natural Science
Foundation of China (11275115), Shandong Province Natural Science Foundation (ZR2010AM023), SRF for ROCS, SEM, and Independent Innovation Foundation of Shandong University }}

\author{%
      ZHU Wei$^{1}$
\quad LIU Yan-Rui$^{1;1)}$\email{yrliu@sdu.edu.cn}%
\quad YAO Tao$^{1}$
}
\maketitle

\address{%
$^1$ School of Physics, Shandong University, Jinan, Shandong 250100, China\\
}

\begin{abstract}
The confirmation of charged charmonium-like states indiates that heavy quark molecules should exist. We here discuss the possibility of a molecule state with $J^{PC}=3^{-+}$. In a one-boson-exchange model investigation for the S wave $C=+$ $D^*\bar{D}_2^*$ states, one finds that the strongest attraction is in the case $J=3$ and $I=0$ for both $\pi$ and $\sigma$ exchanges. Numerical analysis indicates that this hadronic bound state might exist if a phenomenological cutoff parameter around 2.3 GeV (1.5 GeV) is reasonable with a dipole (monopole) type form factor in the one-pion-exchange model. The cutoff for binding solutions may be reduced to a smaller value once the $\sigma$ exchange contribution is included. If a state around the $D^*\bar{D}_2^*$ threshold ($\approx$4472 MeV) in the channel $J/\psi\omega$ (P wave) is observed, the heavy quark spin symmetry implies that it is not a $c\bar{c}$ meson and the $J^{PC}$ are likely to be $3^{-+}$.
\end{abstract}




\begin{multicols}{2}

\section{Introduction}\label{sec1}

Mesons with exotic properties play an important role in understanding the nature of strong interactions. The observation of the so called XYZ states in the heavy quark sector has triggered lots of discussions on their quark structures, decays, and formation mechanisms. It also motivates people to study new states beyond the quark model assignments.

The X(3872), first observed in the $J/\psi\pi^+\pi^-$ invariant mass distribution by Belle collaboration in 2003 \cite{X3872-belle}, is the strangest heavy quark state. Since its extreme closeness to the $D^0\bar{D}^{*0}$ threshold, lots of discussions about its properties are based on the molecule assumption. However, it is very difficult to identify the X(3872) as a shallow bound state of $D^0\bar{D}^{*0}$ since there are no explicitly exotic molecule properties.

A charged charmonium- or bottomonium-like meson labeled as $Z$ is absolutely exotic because its number of quarks and antiquarks must be four or more. Such states include the $Z(4430)$ observed in the $\psi'\pi^\pm$ mass distribution \cite{Z4430-belle}, the $Z_1(4050)$ and $Z_2(4250)$ observed in the $\chi_{c1}\pi^+$ mass distribution \cite{Z1Z2-belle}, the $Z_b(10610)$ and $Z_b(10650)$ in the mass spectra of the $\Upsilon(nS)\pi^\pm$ ($n$=1,2,3) and $\pi^\pm h_b(mP)$ ($m$=1,2) \cite{Zb-belle}, and charged structures $Z_c(3900)$, $Z_c(3885)$, $Z_c(4020)$, and $Z_c(4025)$ observed by BESIII \cite{charged-bes3}. The $Z_c(3900)$ and $Z(4430)$ have been confirmed by analyses from different data \cite{confirmed}. The existence of multiquark states seems to be true. Since $Z(4430)$ is around the $D^*D_1$ threshold, $Z_b(10610)$ is around the $BB^*$ threshold, $Z_b(10650)$ is around the $B^*B^*$ threshold, and $Z_c(3900)$ is around the $D\bar{D}^*$ threshold, molecular models seem to be applicable to their structure investigations \cite{LLDZ08-4430mole,DingHLY09,SunHLLZ,ZhangZH11,OhkodaYYSH,YangPDZ12,LiWDZ13,WHZ}.

To identify a state as a molecule is an important issue in hadron studies. One should consider not only bound state problem of two hadrons, but also how to observe a molecular state in possible production processes. In Refs. \cite{Nstar-dyn,Nstar-chiqm,Nstar-obe,Nstar-cc}, bound states of $\Sigma_c\bar{D}$ and $\Sigma_c\bar{D}^*$ were studied. Since the quantum numbers are the same as the nucleon but the masses are much higher, identifying them as multiquark baryons is rather apparent. To obtain a deeper understanding of the strong interaction, it is necessary to explore possible molecules with explicitly exotic quantum numbers.

Quark model gives us a constraint on the quantum numbers of a meson, namely, a meson with $J^{PC}=0^{--}$, $0^{+-}$, $1^{-+}$, $2^{+-}$, $3^{-+}$, $\cdots$ could not be a $q\bar{q}$ state, but it may be a multiquark state. So the study on such states may deepen our understanding of nature. If two $q\bar{q}$ mesons can form a molecule with such quantum numbers, one gets the simplest configuration. Next simpler configuration is the baryon-antibaryon case. A possible place to search for them is around hadron-hadron thresholds. There are some discussions on low spin heavy quark exotic states in Refs. \cite{ShenCLHZYL10,HuCLHZYL11}. Here we would like to discuss the possibility of a higher spin state, $J^{PC}=3^{-+}$. One will see that identification of it from strong decay is possible.

First, we check meson-antimeson systems that can form $3^{-+}$ states, where meson (antimeson) means that its quark structure is $c\bar{q}$ ($\bar{c}q$). The established mesons may be found in the Particle Data Book \cite{PDG}. One checks various combinations and finds that the lowest S-wave system is $D^*\bar{D}_2^*$. The next S-wave one is $D_s^*D_{s2}^*$. Between these two thresholds, one needs $D$ or $G$ wave to combine other meson-antimeson pairs (see Fig. \ref{th3-}). Below the threshold of $D^*\bar{D}_2^*$, the orbital angular momentum is $D$, $F$, or $G$-wave. Above the $D_s^*D_{s2}^*$ threshold, a partial wave of $P$, $F$, or $H$ is needed. Since the difference between these two thresholds is more than 200 MeV, one may neglect the channel coupling and choose the $D^*D_2^*$ system to study.

\begin{center}
\includegraphics[width=4cm]{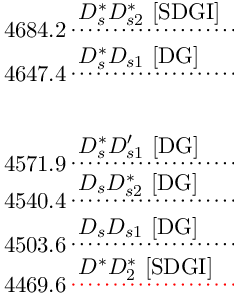}
\figcaption{\label{th3-}Thresholds of $J^{PC}=3^{-+}$ meson-antimeson systems between that of $D^*D_2^*$ and that of $D_s^*D_{s2}^*$. $S$, $D$, $G$, and $I$ are orbital angular momenta.}
\end{center}

Secondly, we check baryon-antibaryon systems. If one combines the established $cqq$ baryons and their antibaryons, one finds that the lowest S-wave threshold is for $\Lambda_c(2880)$ and $\bar{\Lambda}_c$ ($\approx5168$ MeV). Even for the lowest threshold of $\Lambda_c(2595)$ and $\bar{\Lambda}_c$ in F-wave, the value ($\approx4879$ MeV) is still higher than that of $D_s^*D_{s2}^*$. Thus, we may safely ignore the possible baryon-antibaryon contributions in this study.

In a $3^{-+}$ $D^*\bar{D}_2^*$ state, partial waves of $S$, $D$, $G$, and $I$ may all contribute. As a first step exploration, we consider only the dominant S-wave interactions. Possible coupled channel effects will be deferred to future works. The present study is organized as follows. After the introduction in Sec. {\ref{sec1}}, we present the main ingredients for our study in Sec. {\ref{sec2}}. Then we give the numerical results in Sec. {\ref{sec3}}. The final part is for discussions and conclusions.

\section{Wavefunctions, amplitudes, and Lagrangian}\label{sec2}

We study the meson-antimeson bound state problem in a meson exchange model. The potential is derived from the scattering amplitudes \cite{LLDZ08-4430wave} and the flavor wave functions of the system are necessary. Since the states we are discussing have a definite C-parity while the combination of $c\bar{q}$ and $\bar{c}q$ mesons does not, a relative sign problem arises between the two parts of a flavor wave function. One has to find the relation between the flavor wave function and the potential with definite C-parity. There are some discussions about this problem in the literatures \cite{LLDZ08-4430wave,LLDZ08-3872wave,ThomasC08,Stancu08-wave,LiuZ09}. Here we revisit it by using the G-parity transformation rule which relates the amplitudes between $NN$ and $N\bar{N}$ \cite{KlemptBMR}. The final potential is irrelevant with the relative sign.

Since $D$ mesons do not have defined C-parity, we may assume arbitrary complex phases $\alpha$ and $\beta$ under the C-parity transformations
\begin{eqnarray}
&D^{*-}\leftrightarrow \alpha_1D^{*+},\qquad\bar{D}^{*0}\leftrightarrow \alpha_2 D^{*0},&\nonumber\\
&D^{*-}_2\leftrightarrow \beta_1D^{*+}_2,\qquad\bar{D}^{*0}_2\leftrightarrow\beta_2 D^{*0}_2.&
\end{eqnarray}
According to the SU(2) transformation, one finds the following isospin doublets
\begin{eqnarray}
\left(\begin{array}{c}\bar{D}^{*0}\\D^{*-}\end{array}\right),\quad
\left(\begin{array}{c}\alpha_1D^{*+}\\-\alpha_2 D^{*0}\end{array}\right),\nonumber\\
\left(\begin{array}{c}\bar{D}^{*0}_2\\D^{*-}_2\end{array}\right),\quad
\left(\begin{array}{c}\beta_1D^{*+}_2\\-\beta_2 D^{*0}_2\end{array}\right),
\end{eqnarray}
from which the G-parity transformations read
\begin{eqnarray}\label{G-rule}
&\left(\begin{array}{c}\bar{D}^{*0}\\D^{*-}\end{array}\right)\rightarrow \left(\begin{array}{c}\alpha_1D^{*+}\\-\alpha_2 D^{*0}\end{array}\right)\rightarrow\left(\begin{array}{c}-\bar{D}^{*0}\\-D^{*-}\end{array}\right),&\nonumber\\
&\left(\begin{array}{c}\bar{D}^{*0}_2\\D^{*-}_2\end{array}\right)\rightarrow \left(\begin{array}{c}\beta_1D^{*+}_2\\-\beta _2 D^{*0}_2\end{array}\right)\rightarrow\left(\begin{array}{c}-\bar{D}^{*0}_2\\-D^{*-}_2\end{array}\right).&
\end{eqnarray}

Similar to the study of the $D^*\bar{D}_1$ bound state problem \cite{LLDZ08-4430wave}, one may construct several states from $D^*$ and $\bar{D}_2^*$. Here we concentrate only on the $C=+$ case. If the system is an isovector (isoscalar), we label it $Z_J$ ($X_J$) where $J$ is the angular momentum. Explicitly, one has the G-parity eigenstates
\end{multicols}
\begin{eqnarray}
Z_J^0&=&\frac{1}{2\sqrt2}\Big[(D^{*-}D_2^{*+}+(-1)^{J-3}D_2^{*+}D^{*-})-\beta_1^\dag\beta_2(\bar{D}^{*0}D_2^{*0}+(-1)^{J-3}D_2^{*0}\bar{D}^{*0})\nonumber\\
&&+c\alpha_1\beta_1^\dag ((-1)^{J-3}D_2^{*-}D^{*+}+D^{*+}D_2^{*-})-c\alpha_2\beta_1^\dag((-1)^{J-3}\bar{D}_2^{*0}D^{*0}+D^{*0}\bar{D}_2^{*0})\Big],\nonumber
\end{eqnarray}
\begin{eqnarray}
X_J^0&=&\frac{1}{2\sqrt2}\Big[(D^{*-}D_2^{*+}+(-1)^{J-3}D_2^{*+}D^{*-})+\beta_1^\dag\beta_2(\bar{D}^{*0}D_2^{*0}+(-1)^{J-3}D_2^{*0}\bar{D}^{*0})\nonumber\\
&&+c\alpha_1\beta_1^\dag ((-1)^{J-3}D_2^{*-}D^{*+}+D^{*+}D_2^{*-})+c\alpha_2\beta_1^\dag((-1)^{J-3}\bar{D}_2^{*0}D^{*0}+D^{*0}\bar{D}_2^{*0})\Big],
\end{eqnarray}
where $c=1$ is the C-parity and the superscript indicates the electric charge. The factor $(-1)^{J-3}$ is from the exchange of two bosons \cite{Liu13}. One may check
\begin{eqnarray}
\hat{G}Z_J^0=-cZ_J^0,\quad \hat{C}X_J^0=cX_J^0.
\end{eqnarray}

The procedure to derive the potential is similar to that in Ref. \cite{LLDZ08-4430wave}. Now we calculate the amplitude $T(Z_J^0)=\langle Z_J^0|\hat{T}|Z_J^0\rangle$ with the G-parity transformation rule (\ref{G-rule}). We just present several terms to illustrate the derivation. Together with the above $Z_J^0$ wave function, one has
\begin{eqnarray}
T(Z_J^0)&=&\frac14\left\{
T_{[D_2^{*+}\to D_2^{*+}, D^{*-}\to D^{*-}]}-\beta_1\beta_2^\dag T_{[D_2^{*+}\to D_2^{*0}, D^{*-}\to \bar{D}^{*0} ]}+c\alpha_1^\dag \beta_1(-1)^{J-3}T_{[D_2^{*+}\to D^{*+}, D^{*-}\to D_2^{*-}]}+\cdots\right\}\nonumber\\
&=&\frac{G^\pi}{4}\left\{
T_{[D_2^{*+}\to D_2^{*+}, D^{*0}\to D^{*0}]}+\alpha_1^\dag\alpha_2\beta_1\beta_2^\dag T_{[D_2^{*+}\to D_2^{*0}, D^{*0}\to {D}^{*+}]}+c\alpha_1^\dag\alpha_2\beta_1\beta_2^\dag(-1)^{J-3} T_{[D_2^{*+}\to D^{*+}, D^{*0}\to D_2^{*0}]}+\cdots\right\}.\nonumber\\
\end{eqnarray}

In fact, the convention $\alpha_1\alpha_2^\dag=\beta_1\beta_2^\dag$ is implied in the Lagrangian in Eq. (\ref{Lag-pi}). So $\alpha_1\alpha_2^\dag\beta_1^\dag\beta_2=\alpha_1^\dag\alpha_2\beta_1\beta_2^\dag=1$ and one finally gets
\begin{eqnarray}\label{T-matrices}
T_J&=&\frac12G^\pi\left\{
T_{[D_2^{*+}\to D_2^{*+}, D^{*0}\to D^{*0}]}+x T_{[D_2^{*+}\to D_2^{*0}, D^{*0}\to {D}^{*+}]}+x T_{[D_2^{*0}\to D_2^{*+},{D}^{*+}\to D^{*0}]}+T_{[D_2^{*0}\to D_2^{*0},{D}^{*+}\to {D}^{*+}]}\right.\nonumber\\
&&+c(-1)^{J-3} T_{[D_2^{*+}\to D^{*+}, D^{*0}\to D_2^{*0}]}+xc(-1)^{J-3} T_{[D_2^{*+}\to D^{*0}, D^{*0}\to{D}_2^{*+} ]}+xc(-1)^{J-3} T_{[D_2^{*0}\to D^{*+},{D}^{*+}\to D_2^{*0}]}\nonumber\\
&&\left.+c(-1)^{J-3} T_{[D_2^{*0}\to D^{*0},{D}^{*+}\to{D}_2^{*+}]}\right\},
\end{eqnarray}
\begin{multicols}{2}
\noindent where $x=1$ (-1) for $I=1$ (0) state. It is obvious that we may calculate the potential of meson-antimeson interaction from that of meson-meson together with a given Lagrangian for ($c\bar{q}$) meson fields. The arbitrary relative phase in the flavor wave function of a meson-antimeson system is canceled in this procedure. To derive the explicit expression of the potential, one needs interaction Lagrangian.

The Lagrangian for pion interactions in the heavy quark limit and chiral limit reads \cite{hchi-coupling,FalkL92}
\begin{eqnarray}\label{Lag-pi}
\mathcal{L}_\pi&=&g {\rm Tr}[H {A}\!\!\!\slash\gamma_5\bar{H}]
+g''{\rm Tr}[T_{\mu}A\!\!\!\slash\gamma_5\bar{T}^{\mu}]\nonumber\\
&&+\{\frac{h_1}{\Lambda_{\chi}}{\rm Tr}[T^{\mu}(D_{\mu}{A}\!\!\!\slash)\gamma_5\bar{H}]+h.c.\}\nonumber\\
&&+\{\frac{h_2}{\Lambda_{\chi}}{\rm Tr}[T^{\mu}(D\!\!\!\!/A_{\mu})\gamma_5\bar{H}]+h.c.\},
\end{eqnarray}
where
\begin{eqnarray}
H&=&\frac{1+v\!\!\!/}{2 }[P^{*\mu}\gamma_\mu+P \gamma_5],\quad\nonumber\\
T^{\mu}&=&\frac{1+v\!\!\!/}{2}\Big\{P^{*\mu\nu}_{2}
\gamma_{\nu}+\sqrt{\frac{3}{2}}P_{1}^{\nu}\gamma_5 [g_{\nu}^{\mu}-\frac{1}{3}\gamma_{\nu}(\gamma^{\mu}-v^{\mu})]\Big\},\nonumber
\end{eqnarray}
\begin{eqnarray}
\bar{H}=\gamma^0H^\dag\gamma^0,\quad \bar{T}^{\mu}=\gamma^0T^\dag\gamma^0.
\end{eqnarray}
The fields $P^*=(D^{*0},D^{*+})$, and $P_2^*=(D_2^{*0},D_2^{*+})$ annihilate the $c\bar{q}$ mesons. $P$ and $P_1$ have similar form but they are not involved in the following calculation. The axial vector field $A^{\mu}$ is defined as $A^{\mu}=\frac{i}{2}(\xi^{\dag}\partial^{\mu}\xi-\xi\partial^{\mu}\xi^{\dag})$
with $\xi=\exp(i\mathcal{M}/f)$, $f=132$ MeV and
\begin{eqnarray}
\mathcal{M}&=&
\left(\begin{array}{cc}
\frac{\pi^{0}}{\sqrt{2}}&\pi^{+}\\
\pi^{-}&-\frac{\pi^{0}}{\sqrt{2}}
\end{array}\right).
\end{eqnarray}

In one boson exchange models of nuclear force, long-range interaction is controlled by pion exchange while the intermediate interaction mainly results from a phenomenological broad $\sigma$ meson. This scalar meson exchange represents an effective description of the 2$\pi$ contributions. Its contribution can be even approximated by a zero-width scalar exchange with suitably adjusted parameters in the Bonn model \cite{Bonn}. In the Nijmegen model, a broad scalar meson $\epsilon$ is described by a two-pole approximation with the lower pole corresponding to the $\sigma$ \cite{Nijmegen}. In the present study of bound state problem, we use a zero-width approximation for the $\sigma$. In principle, this economical description may capture main feature of the correlated 2$\pi$ contribution. Recent investigations indicate that the pole mass of $\sigma$ is around 400$\sim$600 MeV \cite{sig-mass}. We will use a larger value 600 MeV which provides a weaker attraction. To further consider this $\sigma$ contribution, one needs additional interaction terms
\begin{eqnarray}\label{Lag-sig}
{\cal L}_\sigma&=&g_\sigma{\rm Tr}[H\sigma\bar{H}]+g_\sigma''{\rm Tr}[T^\mu\sigma\bar{T}_\mu]+\frac{h_\sigma'}{f_\pi}{\rm Tr}[T^\mu(\partial_\mu\sigma)\bar{H}\nonumber\\
&&+H(\partial_\mu\sigma)\bar{T}^\mu].
\end{eqnarray}

The coupling constants must be determined in order for numerical analysis. One extracts the pion coupling constant $g$ from the decay $D^*\to D\pi$: $g=0.59\pm0.07\pm0.01$ \cite{g-coupling}. For $h_\chi=\frac{h_1+h_2}{\Lambda_\chi}$, we use the value $0.55\text{ GeV}^{-1}$ estimated in Ref. \cite{hchi-coupling}. To determine the coupling constant $g''$, we turn to the chiral quark model \cite{ZhangYSDFS97} with which one may get the relation $g''=-g$.

For the sigma coupling constants, we can just get estimates from the chiral quark model or the chiral multiplet assumption \cite{BardeenEH03}. These approaches have been used in the baryon case \cite{LiuO12} for the purpose of cross checking, where we get consistent results. Now the former method may give the relation $g_\sigma''=-g_\sigma$ and the value $g_\sigma=g_{ch}=2.621$ if one adopts the Lagrangian \cite{ZhangYSDFS97}
\begin{eqnarray}
L_I&=&-g_{ch}\bar{\psi}(\sigma+i\gamma_5\pi_a\tau_a)\psi,
\end{eqnarray}
where $\psi=(u,d)^T$ is the quark field and $\tau_a$ the Pauli matrix. One should note the normalization problem in this approach \cite{YanCCLLY92,FalkL92}. However, if one estimates $g_\sigma$ from the chiral multiple assumption, a value less than 1 is obtained. For the remaining $h_\sigma'$, no available approach may be used. Since the large uncertainties of the coupling constants, we will select several values to see the $\sigma$-exchange effects on the conclusions.

In deriving the above relations for the coupling constants, we have used the polarization vectors $\varepsilon_{\pm1}^\mu=\frac{1}{\sqrt2}(0,\pm1,i,0)$ and $\varepsilon_{0}^\mu=(0,0,0,-1)$ for the vector meson $D^*$ and
\begin{eqnarray}
\varepsilon_{\pm2}^{\mu\nu}&=&\varepsilon_{\pm1}^\mu\varepsilon_{\pm1}^\nu,\nonumber\\
\varepsilon_{\pm1}^{\mu\nu}&=&\sqrt{\frac12}[\varepsilon_{\pm1}^\mu\varepsilon_0^\nu+\varepsilon_0^\mu\varepsilon_{\pm1}^\nu],\nonumber\\
\varepsilon_0^{\mu\nu}&=&\sqrt{\frac16}[\varepsilon_{+1}^\mu\varepsilon_{-1}^\nu+\varepsilon_{-1}^\mu\varepsilon_{+1}^\nu+2\varepsilon_0^\mu\varepsilon_0^\nu],
\end{eqnarray}
for the tensor meson $D_2^*$ \cite{ChengY11}.

\section{Potentials and numerical analysis}\label{sec3}

Now one may derive the potentials through the amplitudes in (\ref{T-matrices}). Using the same procedure as Ref. \cite{LLDZ08-4430wave}, one gets the one-pion-exchange potential (OPEP) for S-wave interaction in the case $I=1$
\end{multicols}
\begin{eqnarray}
V^\pi(Z_J)&=&-\frac{gg''}{6f^2}G^\pi C_d\Big[\delta(\vec{r})-\frac{m_\pi^2e^{-m_\pi r}}{4\pi r}\Big]
+\frac{|h_\chi|^2}{15f^2}G^\pi c(-1)^{J-3} C_e\Big[\nabla^2\delta(\vec{r})-\mu^2\delta(\vec{r})-\frac{\mu^4}{4\pi r}\cos(\mu r)\Big],
\end{eqnarray}
where $\mu=\sqrt{(m_{D_2}-m_{D^*})^2-m_\pi^2}$, and
\begin{eqnarray}
C_d&=&\left\{\begin{array}{rl}
-1,&(J=3)\\
\frac{1}{2},&(J=2)\\
\frac{3}{2},&(J=1)
\end{array}\right.,
\quad
C_e=\left\{\begin{array}{rl}
\frac{1}{2},&(J=3)\\
-\frac{5}{4},&(J=2)\\
-\frac{3}{4},&(J=1)
\end{array}\right..
\end{eqnarray}
There are two parts in the potential: direct part and spin-exchange part. The later corresponds to the terms containing $c$ in Eq. (\ref{T-matrices}). For the case of $I=0$, $V^\pi(X_J)=-3V^\pi(Z_J)$.

The singular behavior at small distances needs to be regularized \cite{Tornqvist94}. If a form factor $FF=\left(\frac{\Lambda^2-m^2}{\Lambda^2-q^2}\right)^2$ is added to each vertex, one finally has
\begin{eqnarray}
V^\pi(Z_J)&=&-\frac{gg''}{6f^2}G^\pi C_d
\Big[-\frac{m_\pi ^2}{4\pi r}(e^{-m_\pi r}-e^{-\Lambda r})+\frac{m_\pi ^2\eta^2}{8\pi\Lambda}e^{-\Lambda r}
+\frac{m_\pi^2\eta^4}{32\pi\Lambda^3}(1+\Lambda r)e^{-\Lambda r}
+\frac{\eta^6}{192\pi\Lambda^3}(3+3\Lambda r+\Lambda^2r^2)e^{-\Lambda r}\Big]\nonumber\\
&&+\frac{|h_\chi|^2}{15 f^2}G^\pi c(-1)^{J-3} C_e\Big\{-\frac{\mu^4}{4\pi r}(\cos(\mu r)-e^{-\alpha r})
+\frac{\mu^4\eta^2}{8\pi\alpha}e^{-\alpha r}\nonumber\\
&&-\frac{\mu^2\eta^4}{32\pi\alpha}(1+\alpha r)
e^{-\alpha r}-\frac{\eta^6}{192\pi\alpha}(3+3\alpha r-\alpha^2r^2)e^{-\alpha r}\Big\},
\end{eqnarray}
where $\eta=\sqrt{\Lambda^2-m_\pi^2}$, and $\alpha=\sqrt{\Lambda^2-(m_{D_2}-m_{D^*})^2}$.

Similarly, the one-$\sigma$-exchange potential (OsEP) is
\begin{eqnarray}
V^\sigma(Z_J)&=&g_\sigma g_\sigma''\Big[\frac{1}{4\pi r}(e^{-m_\sigma r}-e^{-\Lambda r})-\frac{\eta_\sigma^2}{8\pi \Lambda}e^{-\Lambda r}
-\frac{\eta_\sigma^4}{32\pi\Lambda^3}(1+\Lambda r)e^{-\Lambda r}\nonumber\\
&&-\frac{\eta_\sigma^6}{192\pi\Lambda^5}(3+3\Lambda r+\Lambda^2r^2)e^{-\Lambda r}\Big]\nonumber\\
&&+\frac{|h_\sigma'|^2}{3f_\pi^2}c(-1)^{J-3}C_\sigma\Big[\frac{\mu_\sigma^2}{4\pi r}(e^{-\mu_\sigma r}-e^{-\alpha r})-\frac{\mu_\sigma^2\eta_\sigma^2}{8\pi\alpha}e^{-\alpha r}
    -\frac{\mu_\sigma^2\eta_\sigma^4}{32\pi\alpha^3}(1+\alpha r)e^{-\alpha r}\nonumber\\
&&-\frac{\eta_\sigma^6}{192\pi\alpha^3}(3+3\alpha r+\alpha^2r^2)e^{-\alpha r}\Big],\nonumber\\
V^\sigma(X_J)&=&V^\sigma(Z_J),
\end{eqnarray}
\begin{multicols}{2}
\noindent where $\mu_\sigma=\sqrt{m_\sigma^2-(m_{D_2}-m_{D*})^2}$, $\eta_\sigma=\sqrt{\Lambda^2-m_\sigma^2}$. The coefficient $C_\sigma=1$ for $J=3$, $\frac12$ for $J=2$, and $\frac16$ for $J=1$. The spin-dependent nature of OsEP comes from the third coupling term in the Lagrangian (\ref{Lag-sig}).

Before the numerical calculation, we take a look at the relative strengthes of potentials. For the meson masses, we use $m_\pi=137.27$ MeV, $m_{D*}=2008.63$ MeV, and $m_{D_2}=2463.5$ MeV \cite{PDG}. We plot OPEPs with $\Lambda=1$ GeV in Fig. \ref{potpi}. It is obvious that $X_3$ is the most attractive case.
\begin{center}
\scalebox{0.5}{\includegraphics{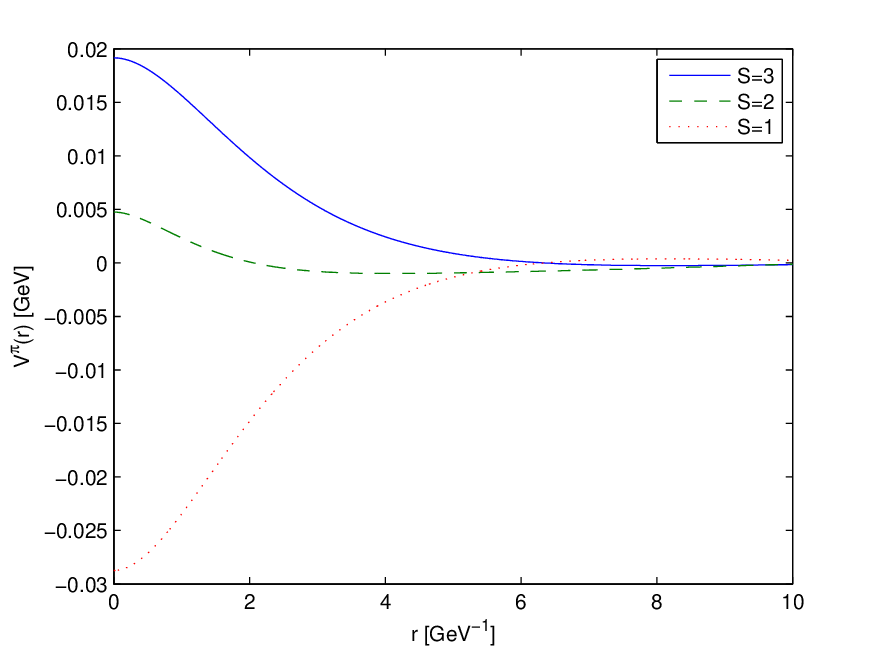}}\\
(a)\\
\scalebox{0.5}{\includegraphics{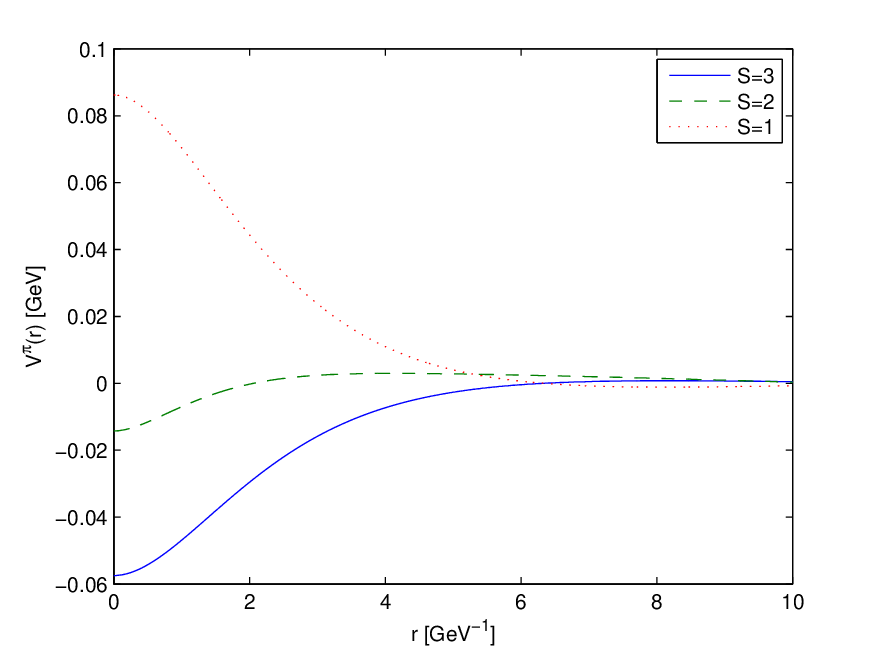}}\\
(b)
\figcaption{One-pion-exchange potentials for (a) Z states and (b) X states with the cutoff $\Lambda=1$ GeV.}\label{potpi}
\end{center}

\begin{center}
\scalebox{0.5}{\includegraphics{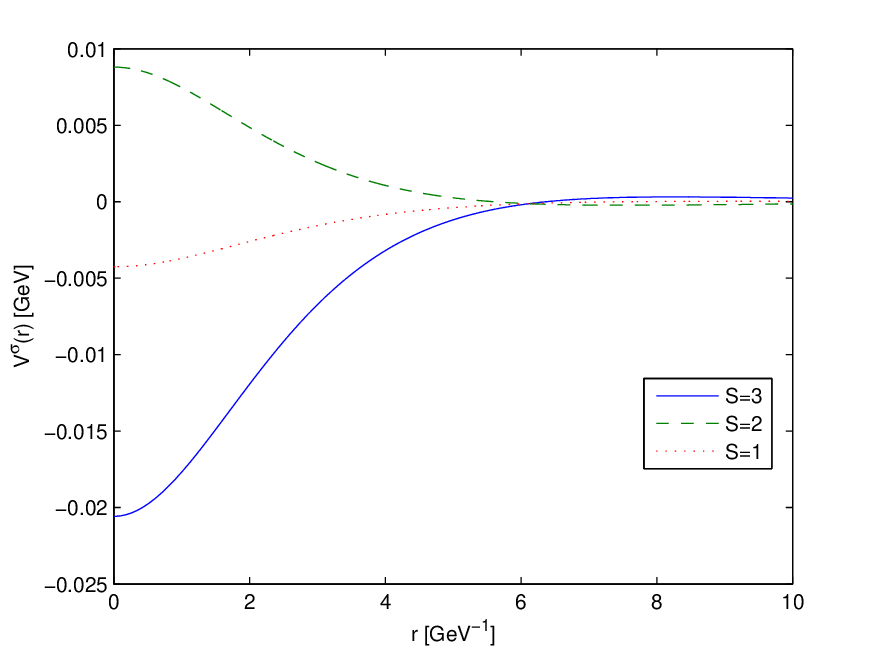}}
\figcaption{One-sigma-exchange potentials for X and Z states with the cutoff $\Lambda=1$ GeV and the coupling constants $g_\sigma''=-g_\sigma=-1.0$, $h_\sigma'=1.0$.}\label{potsig}
\end{center}

In Fig. \ref{potsig}, we show OsEPs with $g_\sigma''=-g_\sigma=-1.0$, $h_\sigma'=1.0$, and $\Lambda=1$ GeV. It is interesting that the potential for $X_3$ is also the most attractive one. Thus the long-range and medium-range meson-exchanges are both helpful for the formation of a $I^G(J^{PC})=0^+(3^{-+})$ state.

Now we turn to the numerical results for the OPEP case by solving the Schr\"odinger equation. In the potential, there is an unknown phenomenological cutoff parameter $\Lambda$. It incorporates the size information of the interacting mesons. If $\Lambda$ goes to infinity, the potential describes the interactions of structureless mesons. A small cutoff is relevant to the real case. In principle, an appropriate value should be around 1 GeV which is realized from the nuclear potential models \cite{Bonn,Julich}. There, the values of cutoffs can be determined by fitting plenty of scattering data. Since the system we are discussing is completely different and no experimental data are available, we just tune the cutoff value and check whether a bound state exists or not if it falls into a reasonable range. The results are sensitive to the cutoff parameter and we tend to use some criteria to constrain the range. Noticing a hadronic molecule is not a tightly bound state and the cutoff can reach 3.0 GeV in the CD-Bonn model \cite{CDBonn}, we will not show results if the root-mean-square radius $r_{rms}<$0.8 fm or $\Lambda>$4 GeV. Binding energy (B.E.) and $r_{rms}$ for $X_3$ with various $\Lambda$ are presented in Tab. \ref{BEpi}. Similarly, one may get numerical results for other possibilities, which are also given in that table. The resultant cutoff much larger than 1 GeV indicates that the attraction is not strong enough for the formation of a hadronic bound state. From the numerical calculations, one does not find binding solutions for $X_1$, $Z_3$, and $Z_2$ with the above criteria. Binding solution exists for $X_2$ in a very narrow range ($\Lambda\sim3.15$ MeV). Thus we give one more significant figure for the cutoff in the table. Among the three cases in the table, of course $X_3$ is the most likely to be existent. If $\Lambda$ around 2.3 GeV is a reasonable value in the OPEP model, the existence of $X_3$ is possible. However, the sensitivity of results to the cutoff does not allow us to reach a definite conclusion.
\begin{center}
\tabcaption{Cutoff ($\Lambda$), binding energy (B.E.) and root-mean-square radius ($r_{rms}$) for $X$ and $Z$ states with OPEP. We do not show results if $\Lambda>4$ GeV or $r_{rms}<0.8$ fm. We present one more significant figure for the cutoff if the results are very sensitive to it.}\label{BEpi}
\footnotesize
\begin{tabular*}{80mm}{c@{\extracolsep{\fill}}ccc}
\toprule State&$\Lambda$ (GeV)& B.E. (MeV) & $r_{rms}$ (fm)\\
$X_3$&2.3 &0.6&3.8\\
&2.4 &3.7&1.6\\
&2.5 &9.9&1.0\\
&2.6 &19.8&0.8\\\hline
$X_2$&3.13&1.0&2.8\\
&3.14&3.5&1.5\\
&3.15&7.0&1.1\\
&3.16&11.5&0.8\\\hline
$Z_1$&3.6 &1.9&2.2\\
&3.7 &8.4&1.1\\
\bottomrule
\end{tabular*}
\end{center}

\begin{center}
\tabcaption{Cutoff values (GeV) for $X$ and $Z$ states with OPEP+OsEP when binding solutions exist. We do not show cutoffs if $\Lambda>4$ GeV or $r_{rms}<0.8$ fm. We present one more significant figure for the cutoff if the results are very sensitive to it.}\label{Cuts-ps}
\footnotesize
\begin{tabular*}{80mm}{c@{\extracolsep{\fill}}ccc}
\toprule States& Set 1 & Set 2 & Set 3\\
$X_3$ &1.7$\sim$2.2&1.5$\sim$1.7&1.0$\sim$1.1\\
$X_2$ &2.8$\sim$2.9&3.63$\sim$3.64&$>$3.5\\
$X_1$ &&&\\
$Z_3$ &&2.7$\sim$3.0&1.1$\sim$1.2\\
$Z_2$ &&&$>$3.3\\
$Z_1$ &2.2$\sim$2.9&2.8$\sim$3.1&1.5$\sim$1.7\\
\bottomrule
\end{tabular*}
\end{center}

The minimal cutoff for a binding solution is a little larger than 2 GeV if we consider only $\pi$-exchange. One may understand that other contributions have been encoded in the cutoff parameter in the OPEP model. This means that additional attraction may lower the value to a more appropriate number. We would like to check how much attraction the sigma meson contributes. Because of the large uncertainty for the coupling constants, we take three sets of them: (1) $g_\sigma$=2.621, $g_\sigma''=-g_\sigma$, $h_\sigma'=0$; (2) $g_\sigma=1.0$, $g_\sigma''=-g_\sigma$, $h_\sigma'=1$; and (3) $g_\sigma=2.621$, $g_\sigma''=-g_\sigma$, $h_\sigma'=2.621$. After the solution of the Schr\"odinger equation, the cutoff parameters satisfying the condition $r_{rms}>$0.8 fm and $\Lambda<$4 GeV are summarized in Tab. \ref{Cuts-ps}. Set 1 corresponds to the case without spin-exchange sigma potential. In this case, the existence of $X_3$ is possible if the cutoff around 1.7 GeV is a reasonable value. Set 2 has a larger spin-dependent sigma potential and a smaller spin-independent sigma potential. Conclusion in this case is similar to set 1. The last set is the most attractive case, where the arbitrary number for the $h^\prime_\sigma$ might be a large value. In this case, $X_3$, $Z_3$, and $Z_1$ all seem to be existent. Comparing set 1 and set 3, one sees that the spin-exchange sigma potential may give an important contribution. From the comparison with the OPEP case, one sees that the cutoff value reduced by the sigma meson exchange depends a lot on the unfixed coupling constants. The readers may draw their own conclusions for the importance of the sigma contributions from Tab. \ref{Cuts-ps}. If the bound state $X_3$ really exists, the structure should be observed around the $D^*D_2^*$ threshold ($\approx4472$ MeV).

In the above investigation, we have added a dipole type form factor at each vertex to regularize the original potential. One may also use a monopole type form factor $FF=\Big(\frac{\Lambda^2-m^2}{\Lambda^2-q^2}\Big)$. Now the obtained potentials are
\end{multicols}
\begin{eqnarray}
V^\pi(Z_J)&=&-\frac{gg''}{6f^2}G^\pi C_d
\Big[-\frac{m_\pi ^2}{4\pi r}(e^{-m_\pi r}-e^{-\Lambda r})+\frac{\Lambda\eta^2}{8\pi}e^{-\Lambda r}\Big]\nonumber\\
&&+\frac{|h_\chi|^2}{15 f^2}G^\pi c{(-1)^{J-3}} C_e\Big\{-\frac{1}{4\pi r}\Big[\mu^4\cos(\mu r)-\alpha^4e^{-\alpha r}+2\alpha^2\eta^2e^{-\alpha r}\Big]
+\frac{\alpha^3\eta^2}{8\pi}e^{-\alpha r}\Big\},\\
V^\sigma(Z_J)&=&g_\sigma g_\sigma''\Big[\frac{1}{4\pi r}(e^{-m_\sigma r}-e^{-\Lambda r})-\frac{\eta_\sigma^2}{8\pi \Lambda}e^{-\Lambda r}\Big]\nonumber\\
&&+\frac{|h_\sigma'|^2}{3f_\pi^2}{c(-1)^{J-3}}C_\sigma\Big[\frac{\mu_\sigma^2}{4\pi r}(e^{-\mu_\sigma r}-e^{-\alpha r})-\frac{\alpha\eta_\sigma^2}{8\pi}e^{-\alpha r}\Big].
\end{eqnarray}
\begin{multicols}{2}
These functions are simpler than the previous ones. However, the resultant cutoff satisfying our criteria is now smaller (see results in Tab. \ref{BEpiM} and \ref{Cuts-psM}). The results are more sensitive to the cutoff parameter. One may understand the feature from the differences in the regularization. Since the original potential has a second order derivative term on the delta function, the regularized potential is not finite at origin in the monopole case while it is finite in the dipole case. Therefore, the singular behavior of the potential in the monopole case is not purely regularized and the sensitivity to the cutoff is higher. The relation between the two cutoffs in the nuclear case is around $\Lambda_{dipole}\approx\sqrt{2}\Lambda_{monopole}$ \cite{Bonn}. Here and in Ref. \cite{DDbar}, we also observe $\Lambda_{dipole}>\Lambda_{monopole}$ for similar binding solutions between these two cases. In the monopole case of OPEP model, if cutoff around 1.5 GeV is reasonable, one gets a possible $X_3$ bound state. The existence of $X_2$ is also possible. In the OPEP+OsEP model, $X_3$ (also $Z_3$ and $Z_1$) is possible for a cutoff around 1.2 GeV.

\begin{center}
\tabcaption{Cutoff ($\Lambda$), binding energy (B.E.) and root-mean-square radius ($r_{rms}$) for $X$ and $Z$ states with OPEP and monopole FF. We do not show results if $\Lambda>4$ GeV or $r_{rms}<0.8$ fm. We present one more significant figure for the cutoff if the results are very sensitive to it.}\label{BEpiM}
\footnotesize
\begin{tabular*}{80mm}{c@{\extracolsep{\fill}}ccc}
\toprule State&$\Lambda$ (GeV)& B.E. (MeV) & $r_{rms}$ (fm)\\
$X_3$&1.5 &4.5&1.5\\\hline
$X_2$&1.66&1.2&2.4\\
&1.67&9.9&0.9\\\hline
$Z_1$&2.2 &6.2&1.2\\
\bottomrule
\end{tabular*}
\end{center}

\begin{center}
\tabcaption{Cutoff values (GeV) for $X$ and $Z$ states with OPEP+OsEP and monopole FF when binding solutions exist. We do not show cutoffs if $\Lambda>4$ GeV or $r_{rms}<0.8$ fm. We present one more significant figure for the cutoff if the results are very sensitive to it.}\label{Cuts-psM}
\footnotesize
\begin{tabular*}{80mm}{c@{\extracolsep{\fill}}ccc}
\toprule States& Set 1 & Set 2 & Set 3\\
$X_3$ &1.2$\sim$1.4&1.0$\sim$1.1&$\sim$0.8\\
$X_2$ &$\sim$1.6&$\sim$1.8&$\sim$2.23\\
$X_1$ &&&2.7$\sim$3.1\\
$Z_3$ &&1.8$\sim$2.0&0.81$\sim$0.87\\
$Z_2$ &3.0$\sim$3.8&&$>$2.9\\
$Z_1$ &1.5$\sim$1.9&1.8$\sim$1.9&1.1$\sim$1.2\\
\bottomrule
\end{tabular*}
\end{center}

\section{Discussions and conclusions}\label{sec4}

From the meson exchange potentials, one has found that the most attractive one appearing in the $C=+$ $D^*\bar{D}_2^*$ system is for $X_3$. In the OPEP model, the S-wave molecule $X_3$ is possible if the cutoff parameter around 2.3 GeV (1.5 GeV) is reasonable with a dipole (monopole) form factor introduced at each vertex. In the OPEP+OsEP model, a lower cutoff around 1.7 GeV (1.2 GeV) may result in the binding solution for $X_3$. Probably the reason for the sensitivity to the cutoff parameter is the incompleteness in considering balances among various contributions to the molecule problem. Whether a bound $X_3$ state exists or not needs more elaborate investigations. For example, higher partial waves and channel coupling effects also have contributions to the $3^{-+}$ state, which might afford additional attraction. However, the width of the considered $\sigma$ meson, the decays of the charmed components, and additional meson exchanges might reduce the attraction. Future studies on such effects may be helpful for the understanding on exotic states.

In one boson exchange models of nuclear forces, short-range vector meson exchanges provide strong repulsive force. In the study of a meson-antimeson bound state problem, the contributions from the $\rho$ and $\omega$ exchanges may also be important. However, the inclusion of them introduces two more coupling constants which could not be determined reliably at present. One has noticed that the large uncertainty for the sigma meson coupling constant results in the difficulty in drawing a conclusion. The inclusion of vector meson contributions increases the difficulty further. We tend to consider them when coupling constants could be determined in a more reliable way. The tensor force contributions and coupled channel effects may also be important for the bound state problem. The consideration of such effects for the present system needs an improved formalism and we will discuss the effects in a separate work.

If this state really exists, it may decay through its components, i.e. $D^*\to D\pi$, $D_2^*\to D\pi$, or $D_2^*\to D^*\pi$. The $X_3$ may also decay through the quark rearrangement, i.e. the final states are a $c\bar{c}$ meson and a $q\bar{q}$ ($q=u,d$) meson. The later type decay may be used to identify the exotic quantum numbers. Here we focus only on this case.

For convenience of discussion, we assume that $L$ is the relative orbital momentum between the $c\bar{c}$ and the $q\bar{q}$ mesons and relax the isospin requirement temporarily. Since the spins of the charm quark and the light quark in both $D^*$ and $D_2^*$ are parallel, the spin of $c\bar{c}$ in $X_3$ must be 1. According to the heavy quark spin symmetry, the spin of the final charmonium after rearrangement should also be $S=1$. Thus the final $c\bar{c}$ state can only be $\psi$ or $\chi_{cJ}$. The decay channels are obtained as follows:

(1) If the final $c\bar{c}$ is $J/\psi$, the $J^{PC}$ of the produced $q\bar{q}$ meson may be $(1\sim5)^{--}$ for $L=1$, $(1,3,5)^{+-}$ for $L=2$, $(1\sim7)^{--}$ for $L=3$, and so on. After some inspections on the meson masses, one finds that kinematically allowed decays for the $X_3$ molecule are just $J/\psi\rho$ and $J/\psi\omega$ with $L=1,3,5$, and $J/\psi h_1(1170)$ and $J/\psi b_1(1235)$ with $L=2,4$.

If it is $\psi(2S)$, the kinematically allowed decays are $\psi(2S)\rho$ and $\psi(2S)\omega$ with $L=1,3,5$.

(2) If the $c\bar{c}$ is $\chi_{c0}$, the $J^{PC}$ of the $q\bar{q}$ meson may be $(2\sim4)^{++}$ for $L=1$, $(2,4)^{-+}$ for $L=2$, $(0\sim6)^{++}$ for $L=3$, and so on. The kinematically allowed decays are $\chi_{c0}f_0(500)$, $\chi_{c0}f_0(980)$, and $\chi_{c0}a_0(980)$ with $L=3$.

(3) If the $c\bar{c}$ is $\chi_{c1}$, the $J^{PC}$ of the $q\bar{q}$ meson may be $(2,4)^{-+}$ for $L=0$, $(1\sim5)^{++}$ for $L=1$, $(0,2,4,6)^{-+}$ for $L=2$, $(0\sim7)^{++}$ for $L=3$, and so on. The kinematically allowed decays are $\chi_{c1}\pi$, $\chi_{c1}\eta$, $\chi_{c1}\eta'$ with $L=2,4$, and $\chi_{c1}f_0(500)$ with $L=3$.

(4) If the $c\bar{c}$ is $\chi_{c2}$, the $J^{PC}$ of the $q\bar{q}$ meson may be $(2,4)^{-+}$ for $L=0$, $(0\sim6)^{++}$ for $L=1$, $(0,2,4,6)^{-+}$ for $L=2$, $(0\sim8)^{++}$ for $L=3$, and so on. The kinematically allowed decays are $\chi_{c2}f_0(500)$ with $L=1,3,5$, and $\chi_{c2}\pi$ and $\chi_{c2}\eta$ with $L=2,4$.

Therefore, the allowed two-body strong decays for $X_3$ are $J/\psi\omega$ (PFH), $\psi(2S)\omega$ (PFH), $J/\psi h_1(1170)$ (DG), $\chi_{c0}f_0(500)$ (F), $\chi_{c0}f_0(980)$ (F), $\chi_{c1}\eta$ (DG), $\chi_{c1}\eta'$ (DG), $\chi_{c1}f_0(500)$ (F), $\chi_{c2}f_0(500)$ (PFH), and $\chi_{c2}\eta$ (DG). There is no S-wave decay. Because high $L$ processes are suppressed and $\psi(2S)$ and $\chi_{c2}$ are excited states, the simplest way to identify $X_3$ may be through the $J/\psi\omega$ channel.

Let us analyze the $J^{PC}$ of an assumed state $X(4472)$ observed in the $J/\psi\omega$ mass distribution. Since $J/\psi$ and $\omega$ are both $J^{PC}=1^{--}$ mesons, the quantum numbers of $J/\psi\omega$ are $(0,1,2)^{++}$ for S-wave combination, $(0\sim3)^{-+}$ for P-wave combination, $(0\sim4)^{++}$ for D-wave combination, and so on. What we are interested in is the case that the partial wave is determined to be $P$. If $X$ were a conventional $c\bar{c}$ meson, the state is $\eta_c(4472)$ and the spin of $c\bar{c}$ must be 0. Because of the heavy quark spin symmetry, the decay $\eta_c(4472)\to J/\psi\omega$ is suppressed. Then $X(4472)$ could be a hadronic state. Although other meson-antimeson pairs may also form molecules with $J^{PC}=(0\sim2)^{-+}$, the masses are smaller. Therefore, based on our numerical analysis, this $X$ around the $D^*\bar{D}_2^*$ threshold is very likely to be a state with the exotic $J^{PC}=3^{-+}$.

If one wants to look for $Z_3$, one can use those kinematically allowed decay channels, $J/\psi\rho$ (PFH), $\psi(2S)\rho$ (PFH), $J/\psi b_1(1235)$ (DG), $\chi_{c0}a_0(980)$ (F), $\chi_{c1}\pi$ (DG), and $\chi_{c2}\pi$ (DG). The practical way to identify the $J^{PC}$ is to analyze the partial wave of $J/\psi\rho$. The search is also helpful to test the meson exchange models.

Replacing a $c$ quark with a $b$ quark, one may study the bottom case. Because the production of a hidden bottom molecule $B^*\bar{B}_2^*$ needs much higher energy and the production cross section is smaller, it is difficult for experimentalists to explore this case in near future. However, with the replacement $c\to s$, one may study whether there is a bound state or resonance with $J^{PC}=3^{-+}$ near the $K^*K_2(\approx 2322 \text{ MeV})$ threshold. If such a state exists, it may decay into $\omega\phi$ and can be detected.

In short summary, we have investigated whether hadronic bound states exist in the $D^*\bar{D}_2^*$ system in a one-boson-exchange model. The $C=+$ case is discussed in this paper. We find that the $I^G(J^{PC})=0^+(3^{-+})$ $X_3$ state has the most attractive potential. Whether a bound state exists or not depends strongly on a phenomenological cutoff parameter, which we do not have available data to determine. If a value around 2.3 GeV (1.5 GeV) in the one-pion-exchange potential is reasonable for a dipole (monopole) form factor, the bound state is possible. If the molecule really exists, a feasible place to identify it may be in the invariant mass distribution of $J/\psi\omega$ around 4472 MeV. A similar study for a state around 2322 MeV in the $\omega\phi$ mass distribution is also called for.

\end{multicols}

\vspace{-1mm}
\centerline{\rule{80mm}{0.1pt}}
\vspace{2mm}

\begin{multicols}{2}

\end{multicols}

\clearpage

\end{document}